\documentclass[prl,showpacs,amsmath,amssymb,twocolumn, 9pt]{revtex4}

\usepackage{amsthm}
\usepackage{dcolumn}
\usepackage{bm}
\usepackage{graphicx}

\begin{document}

\newtheorem{corollary}{Corollary}
\newtheorem{definition}{Definition}
\newtheorem{example}{Example}
\newtheorem{lemma}{Lemma}
\newtheorem{proposition}{Proposition}
\newtheorem{theorem}{Theorem}
\newtheorem{fact}{Fact}
\newtheorem{property}{Property}
\newcommand{\bra}[1]{\langle #1|}
\newcommand{\ket}[1]{|#1\rangle}
\newcommand{\braket}[3]{\langle #1|#2|#3\rangle}
\newcommand{\ip}[2]{\langle #1|#2\rangle}
\newcommand{\op}[2]{|#1\rangle \langle #2|}

\newcommand{\tr}{{\rm tr}}
\newcommand {\E } {{\mathcal{E}}}
\newcommand {\F } {{\mathcal{F}}}
\newcommand {\diag } {{\rm diag}}

\title{\Large {\bf Separability of Bosonic Systems}}
\author{Nengkun Yu$^{1,2}$}
\email{nengkunyu@gmail.com}
\affiliation{$^1$Institute for Quantum Computing, University of Waterloo, Waterloo, Ontario, Canada\protect\\
$^2$Department of Mathematics $\&$ Statistics, University of Guelph, Guelph, Ontario, Canada}

\begin{abstract}
In this paper, we study the separability of quantum states in bosonic system. Our main tool here is the \lq \lq separability witnesses", and a connection between \lq \lq separability witnesses" and a new kind of positivity of matrices---\lq \lq Power Positive Matrices" is drawn. Such connection is employed to demonstrate that multi-qubit quantum states with Dicke states being its eigenvectors is separable if and only if two related Hankel matrices are positive semidefinite. By employing this criterion, we are able to show that such state is separable if and only if it's partial transpose is non-negative, which confirms the conjecture in [Wolfe, Yelin, Phys. Rev. Lett. (2014)]. Then, we present a class of bosonic states in $d\otimes d$ system such that for general $d$, determine its separability is NP-hard although verifiable conditions for separability is easily derived in case $d=3,4$.
\end{abstract}


\maketitle
\textit{Introduction---}Entanglement, first recognized as a "spooky" feature of quantum machinery by Einstein, Podolsky, and
Rosen \cite{EPR35}, lies at the heart of quantum mechanics. It has been discovered that entanglement
plays an essential role in various fundamental applications and protocols in quantum information science, such as quantum
teleportation, superdense coding and cryptography \cite{BW92,BBCJ+93,BB84}. Moreover, high order of multipartite entanglement has been shown to be requisite to reach
the maximal sensitivity in metrological tasks \cite{GZN+10}.

In multipartite systems, a quantum state is called separable if it can be written as a statistical mixture of product states, otherwise, it is entangled. Research on separability criteria, that is, on computational
methods to determine whether a given state is
separable or entangled, turns out to be a a cumbersome problem and essential subject in quantum
information theory. Starting from the famous PPT
(Positive Partial Transpose) criterion \cite{PER96}, a considerable
number of different separability criterions have been discovered
(see the references in \cite{HHHH09,IOA07}). One fundamental tool of detecting entanglement is entanglement witnesses \cite{HHH96,TER00}, which is equivalent to the method of positive, but not completely positive maps. Entanglement witnesses are observables that completely
characterize separable states and allow to detect entanglement physically. Their origin stems from Hyperplane separation theorem of geometry: the convex sets can be described by hyperplanes. In particular, a witness is an observable,
which is non-negative for separable states, and it can have a negative expectation value for entangled states.

Despite great efforts and considerable progress have been
made, the physical understanding and mathematical description of
its essential characteristics remain however highly nontrivial
tasks, especially when many-particle systems are analyzed. Moreover, it was shown by Gurvits
\cite{GUR04} that this problem is NP-hard. However, it is still possible to have complete criterion for the separability of some interesting certain situations. A problem of great interest is to study the entanglement of bosonic system \cite{ESBL02,TG09,TAH+12,CBF+13,WY14}. For $N$-qubit bosonic system, a natural basis is $N$-qubit Dicke states(unormalized) which are defined as,
\begin{equation*}
  \ket{D_{N,n}} \! := P_{\textrm{sym}}\bigl( \ket{0}^{\otimes n} \otimes \ket{1}^{\otimes N-n} \bigr),
\end{equation*}
with $P_{\textrm{sym}}$ being the projection onto the Bosonic (fully symmetric)
subspace, $i.e.$, $P_{\textrm{sym}} = \frac{1}{N!} \sum_{\pi\in S_N} U_\pi$, the
sum extending over all permutation operators $U_\pi$ of the $N$-qubit systems. It is worth to note that the entanglement of pure Dicke state has been widely studied recently \cite{YU13,DVC00,YCGD10,YGD14,HKWG+09,BKMG+09,BTZLS+09,WKSW09}.

In this Letter, we focus on the problem of the separability criterion for quantum states in bosonic system by considering separable witnesses. We first draw a connection between the separable witnesses of general multi-qubit bosonic states and a new type of positivity of matrices---what we called \lq \lq Power Positivity". This connection is employed to study the separability of $N$-qubit quantum states which being the mixture of Dicke states. In particular, the separable witnesses of such states corresponds to diagonal \lq \lq Power Positive" matrices, that are just polynomials whose value of is always non-negative for non-negative variable. By employing the characterization of non-negative polynomials, an easily evaluated \textit{complete} criterion for the separability of mixture of Dicke states is demonstrated. Moreover, we show that any such separable state can be written as the mixture of $(N+1)(N+2)$ product states. We then study the separability of a class of states whose eigenvectors are generalized $d\otimes d$ Dicke states. It is proved that the separability problem of such states is NP-complete for general $d$, although very simple criterion is demonstrated for $d=3,4$.

\textit{Main Results---} In the $N$-qudit system $\mathcal{H}_1\otimes\mathcal{H}_2\otimes\cdots\otimes\mathcal{H}_N$ with $d$ being the dimension of each Hilbert space $\mathcal{H}_i$, the bosonic space
is a subspace that spanned by pure quantum states which are invariant under the swap of any two subsystems among all $N$ subsystems, $i.e.$, for the swap operator exchanging
the two qudits system $F_{i,j}$,
\begin{equation*}
S:\equiv\{\ket{\psi}:\ket{\psi}=F_{i,j}\ket{\psi},\ \mathrm{for\ all}\ i,\ j\ \mathrm{and\ Swap}\ F\}.
\end{equation*}
A mixed state $\rho$ is called bosonic if its support is a subspace of bosonic space where the support of $\rho$, $Supp(\rho)$, is the subspace spanned by the eigenvectors corresponding to its non-zero eigenvalues. In other words, $\rho=F_{i,j}\rho=\rho F_{i,j}$ holds for any $1\leq i,j\leq N$.

One very simple observation is that, if a bosonic state $\rho$ is separable, $i.e.$, there exist product states (unnormalized) $\otimes_{k=1}^N\op{\alpha_{j_k}}{\alpha_{j_k}}$ such that
\begin{small}
$$\rho=\sum_{j}\bigotimes_{k=1}^N\op{\alpha_{j_k}}{\alpha_{j_k}},$$
\end{small}
then we can choose product states $\ket{\alpha_{j}}^{\otimes N}$, that is
\begin{small}
$$\rho=\sum_{j}\bigotimes_{k=1}^N\op{\alpha_{j}}{\alpha_{j}}.$$
\end{small}
To see this, one only need to observe that
\begin{small}
\begin{equation*}
\bigotimes_{k=1}^N\ket{\alpha_{j_k}}\in S\Rightarrow \exists\ket{\alpha_{j}},\bigotimes_{k=1}^N\ket{\alpha_{j_k}}=\ket{\alpha_{j}}^{\otimes N}.
\end{equation*}
\end{small}
Now, we introduce the separable witnesses for \textit{bosonic} system as a useful tool: For $N$-qudit system, a Hermitian operator $W$ is called a separable witness of bosonic system if $W=P_SWP_S$ with $P_S$ being the projection of the bosonic space $S$ and it satisfies that
\begin{eqnarray*}
  \tr(W\alpha^{\otimes N})\geq 0, \mathrm{for~all}~\alpha=\op{\alpha}{\alpha}.
\end{eqnarray*}
The importance of separability witness is due to the following proposition.
\begin{proposition}
A bosonic state $\rho$ is separable if and only if $\tr(W\rho)\geq 0$ holds for all separability witness $W$ of bosonic system.
\end{proposition}
\textit{Remark}: This proposition can be generalized to the separability of quantum states lying in fixed subspace.

\textit{Proof:}---The only if part simply follows from the above observation about the structure of separable states of bosonic system. To show the validity of the if part, we assume the existence of entangled bosonic state $\rho$ such that $\tr(W\rho)\geq 0$ holds for all separability witness $W$ of bosonic system. Notice that the set of separable states of bosonic system is convex and compact. Entangled $\rho$ does not lie in this set, by hyperplane separation theorem, one can conclude that there exists a $H$ such that $\tr(H\rho)<0$ and $\tr(H\alpha^{\otimes N})\geq 0$ holds for all $\alpha$. Therefore, $\tr(W\alpha^{\otimes N})=\tr(P_SHP_S\alpha^{\otimes N})=\tr(HP_S\alpha^{\otimes N} P_S)=\tr(H\alpha^{\otimes N})\geq 0$ for $W=P_SHP_S$, then $W$ is a separability witness. On the other hand, $\tr(W\rho)=\tr(P_SHP_S\rho)=\tr(HP_S\rho P_S)=\tr(H\rho)<0$, which contradicts to the assumption. \hfill $\blacksquare$

Notice that the set of separable witnesses forms a convex compact set. In order to check the separability of bosonic states, one way is to parameterize the set of separable witnesses, at least the set of extreme points of separable witnesses.

For simplicity, we mainly focus on the separable witnesses of $N$-qubit bosonic system.

Notice that any Hermitian $W=P_SHP_S$ corresponds to a Hermitian matrix $M:=(m_{i,j})_{(N+1)\times(N+1)}$ as follows
\begin{eqnarray*}
W:=\sum_{i,j=0}^N m_{i,j}\op{\widetilde{D_{N,i}}}{\widetilde{D_{N,j}}}
\end{eqnarray*}
where we employe the dual basis of Dicke states as
\begin{equation*}
  \ket{\widetilde{D_{N,n}}}:= {N \choose n}^{-1}P_{\textrm{sym}}\bigl( \ket{0}^{\otimes n} \otimes \ket{1}^{\otimes N-n} \bigr),
\end{equation*}
that is, $\ip{D_{N,m}}{\widetilde{D_{N,n}}}=\delta_{m,n}$.

Now we can derive the condition for $W$ being separable witness: $\tr(W\alpha^{\otimes N})\geq 0$ holds for all one-qubit $\ket{\alpha}$ is equivalent to
\begin{eqnarray*}
&&\tr(W\op{0}{0}^{\otimes N})\geq 0\Leftrightarrow m_{N,N}\geq 0, \\
&&\tr\{W[(\ket{1}+z\ket{0})(\langle{1}|+z^*\langle{0}|)]^{\otimes N}\}\geq 0,
\end{eqnarray*}
for all $z\in \mathbb{C}$. O
ne can observe that the second condition indicates the first one as $|z|\rightarrow \infty$.

Observe that $(\ket{1}+z\ket{0})^{\otimes N}=\sum_{j=0}^N z^j\ket{D_{N,j}}$, we know that the second condition given above is just
\begin{eqnarray*}
\vec{z}^{\dag} M \vec{z} \geq 0~\mathrm{for~all}~\vec{z}=(1,z,z^2,\cdots,z^{N})^{T}\in\mathbb{C}^{N+1}.
\end{eqnarray*}
This is what we called \lq \lq Power Positive Matrix", which is far different from \lq \lq Semi-definite", \lq \lq Complete Positve". Unfortunately, we are not able to give a complete description of the set of \lq \lq Power Positive Matrices", even one can easily conclude that it is a superset of \lq \lq Semi-definite Matrices".

Although it is difficult to check the separability of general $N$-qubit states, we demonstrate an easily verified analytical condition for the separability of the following general diagonal symmetric
states which is necessary and sufficient.

In $N$-qubit bosonic system, one can naturally define the following class of quantum states, so called the general diagonal symmetric states, GDS \cite{WY14},
\begin{equation*}
\rho=\sum_{n=0}^N\chi_n \op{D_{N,n}}{D_{N,n}},
\end{equation*}
where $\chi_n$ represent the eigenvalues in the eigen-decomposition of $\rho$.

Notice that any GDS state $\rho$ enjoys the symmetry that for all diagonal qubit unitary $U_{\theta}=diag\{1,e^{i\theta}\}$,
\begin{equation*}
\rho=U_{\theta}^{\otimes N}\rho~U_{\theta}^{\dag \otimes N}.
\end{equation*}
Thus, for any separability witness $W$, we have
\begin{equation*}
\tr(W\rho)=\tr(WU_{\theta}^{\otimes N}\rho~U_{\theta}^{\dag \otimes N})=\tr(U_{\theta}^{\dag \otimes N}WU_{\theta}^{\otimes N}\rho),
\end{equation*}
$W_0$ is a \lq \lq diagonal" separable witness and $\tr(W_0\rho)=\tr(W\rho)$ with
\begin{equation*}
W_0=\frac{1}{2\pi}\int_{0}^{2\pi}U_{\theta}^{\dag \otimes N}WU_{\theta}^{\otimes N}d\theta=\sum_{k=0}^N m_{k,k}\op{\widetilde{D_{N,k}}}{\widetilde{D_{N,k}}}.
\end{equation*}
Here, a separable witness $W=\sum_{i,j=0}^N m_{i,j}\op{\widetilde{D_{N,i}}}{\widetilde{D_{N,j}}}$ is called diagonal if $m_{i,j}=0$ for $i\neq j$. According to Proposition 1, we know that:
\begin{proposition}
A general diagonal symmetric state $\rho$ is separable if and only if $\tr(W_0\rho)\geq 0$ for all diagonal separable witness $W_0$.
\end{proposition}

Recall the concept of \lq \lq Power Positive Matrix", $W_0$ is a separable witness if and only if $\sum_{k=0}^N m_{k,k}|z|^{2k}$ is always non-negative for all $z\in\mathbb{C}$.
This is equivalent to
\begin{equation*}
g(r)\geq 0~\mathrm{for~all}~r\geq 0,
\end{equation*}
for real coefficient polynomial $g(x):=\sum_{k=0}^N m_{k,k}x^{k}$, whose value $g(x)$ is always non-negative for non-negative $x$.
The characterization of such polynomials is accomplished by the following proposition.
\begin{proposition}
A real coefficient polynomial $g(x)$ satisfies that $g(r)\geq 0$ for all $r\geq 0$ if and only if there exist real coefficient polynomial $P_i(x),Q_i(x)$ such that
\begin{equation*}
g(x)=\sum_i xP_i^2(x)+\sum_iQ_i^2(x).
\end{equation*}
\end{proposition}
\textit{Proof:}---The if part is simple. To show the validity of the only if part, we use the fundamental theorem of algebra,
\begin{equation*}
g(x)=a_0\prod (x-z_k)^{l_k}.
\end{equation*}
For non real root $z_k$, we know that for all real $r$,
\begin{equation*}
(r-z_k)(r-\bar{z_k})=(r-Re(z_k))^2+Im^2(z_k)\geq 0.
\end{equation*}
For non-positive $z_k$, we know that for all $r\geq 0$,
\begin{equation*}
r-z_k=r+(-z_k)\geq 0.
\end{equation*}
For positive $z_k$, its power $l_k$ must be even.

Thus, expanding $g(x)=a_0\prod(x-z_k)^{l_k}$ confirms the only if part. \hfill $\blacksquare$

Invoking the relation between the diagonal separable witness $W_0$ and the polynomial $g(x)$, one can deduce the following,
\begin{proposition}
Extreme point of the diagonal separable witnesses for GDS has one of the following forms
\begin{eqnarray*}
S&=&\sum_{0\leq i,j\leq \frac{N}{2}}a_ia_j\op{\widetilde{D_{N,i+j}}}{\widetilde{D_{N,i+j}}},\\
T&=&\sum_{0\leq i,j\leq \frac{N-1}{2}}b_ib_j\op{\widetilde{D_{N,i+j+1}}}{\widetilde{D_{N,i+j+1}}},
\end{eqnarray*}
with $a_k,b_k\in\mathbb{R}$.
\end{proposition}
Now we are ready to show our main result,
\begin{theorem}
The GDS state $\rho=\sum_{n=0}^N\chi_n \op{D_{N,n}}{D_{N,n}}$ is separable if and only if the following two Hankel Matrices \cite{PAR88} $M_0,M_1$ are positive semi-definite, $i.e.$,
\begin{eqnarray}
M_0:=\left(\begin{array}{cccc}
\chi_0 & \chi_1 & \cdots & \chi_{m_0}\\
\chi_1 & \chi_2 & \cdots & \chi_{m_0+1}\\
\cdots & \cdots & \cdots & \cdots\\
\chi_{m_0} & \chi_{m_0+1} & \cdots & \chi_{2m_0}
\end{array}\right)\geq 0,\\
M_1:=\left(\begin{array}{cccc}
\chi_1 & \chi_2 & \cdots & \chi_{m_1}\\
\chi_2 & \chi_3 & \cdots & \chi_{m_1+1}\\
\cdots & \cdots & \cdots & \cdots\\
\chi_{m_1} & \chi_{m_1+1} & \cdots & \chi_{2m_1-1}
\end{array}\right)\geq 0,
\end{eqnarray}
where $m_0:=[\frac{N}{2}]$ and $m_1:=[\frac{N+1}{2}]$.
\end{theorem}
\textit{Proof:}---According to Proposition 2 and Proposition 4, $\rho$ is separable if and only if $\tr(W_0\rho)\geq 0$ holds for any extreme point $W_0$ of the diagonal separable witnesses for GDS, that is, for all $\vec{a}=(a_0,\cdots,a_{m_0})^T\in\mathbb{R}^{m_0+1}$, $\vec{b}=(b_1,\cdots,b_{m_1})^T\in\mathbb{R}^{m_1}$the following quadratic forms are non-negative,
\begin{eqnarray*}
\tr(S\rho)=\sum_{0\leq i,j\leq m_0}\chi_{i+j}a_ia_j=\vec{a}^TM_0\vec{a}\geq 0,\\
\tr(T\rho)=\sum_{1\leq i,j\leq m_1}\chi_{i+j-1}b_ib_j=\vec{b}^TM_1\vec{b}\geq 0.
\end{eqnarray*}
It is equivalent to the non-negativity of real Hankel Matrices $M_0,M_1$.
\hfill $\blacksquare$

Now we are going to present the rigorous proof of the conjecture from \cite{WY14}.
\begin{theorem}
The GDS state $\rho=\sum_{n=0}^N\chi_n \op{D_{N,n}}{D_{N,n}}$ is separable if and only if it is PPT. More precisely, if and only if it is PPT under the partial transpose of $m_0=[\frac{N}{2}]$ subsystems.
\end{theorem}
\textit{Proof:}---First, it is sufficient to consider the partial transpose of the first $m_0$ subsystems by noticing the symmetric in the bosonic system. Assume $\rho$ is positive under the partial transpose of $m_0=[\frac{N}{2}]$ subsystems, according to Theorem 1, we only need to show $M_0,M_1\geq 0$ of Eq.(1,2).

One can write $\rho^{\Gamma}$ in basis $\ket{D_{m_0,j}}\ket{D_{m_1,k}}$ with $0\leq j\leq m_0,0\leq k\leq m_1$ by verifying the following equations,
\begin{eqnarray*}
\ket{D_{N,n}}&=&\sum_{j=0}^n\ket{D_{m_0,j}}\ket{D_{m_1,n-j}},~\mathrm{for}~n\leq m_0,\\
\ket{D_{N,n}}&=&\sum_{j=n-m_1}^{m_0}\ket{D_{m_0,j}}\ket{D_{m_1,n-j}},~\mathrm{for}~n > m_0,
\end{eqnarray*}
where $m_1=N-m_0$.

Since $\rho^{\Gamma}\geq 0$, then the restriction of $\rho^{\Gamma}$ on subspace spanned by $\{\ket{D_{m_0,j}}\ket{D_{m_1,j}},0\leq j\leq m_0\}$ is still non-negative, direct calculation leads us to the fact that this is just $M_0\geq 0$.

On the other hand, the restriction of $\rho^{\Gamma}$ on subspace spanned by $\{\ket{D_{m_0,j-1}}\ket{D_{m_1,j}},1\leq j\leq m_1\}$ is still non-negative, direct calculation leads us to the fact that this is just $M_1\geq 0$.

Invoking Theorem 1, we can conclude that $\rho$ is separable.
\hfill $\blacksquare$

One can have the following interesting corollary,
\begin{corollary}
GDS state $\rho=\sum_{n=0}^N\chi_n \op{D_{N,n}}{D_{N,n}}$ is positive under the partial transpose of $m_0=[\frac{N}{2}]$ subsystems, then it is positive under the partial transpose of arbitrary subsystems.
\end{corollary}

In the following, we introduce a class of bipartite GDS states, and study the separability of such states: In $d\otimes d$ system, one can define the following general diagonal symmetric states,
\begin{eqnarray*}
\rho=\sum_{i,j=1}^d\chi_{i,j} \op{\psi_{i,j}}{\psi_{i,j}},
\end{eqnarray*}
with $\ket{\psi_{i,j}}:=\begin{cases}\ket{ii} &{\rm if}\ i=j,\\ \ket{ij}+\ket{ji} &{\rm otherwise.}\end{cases}$ being some basis of the bosonic subspace of $d\otimes d$ system, $i.e.$, the symmetric subspace.

Notice that $\rho=(U\otimes U)\rho(U\otimes U)^{\dag}$ holds for all diagonal qudit unitary $U$. Then, $\rho$ is separable if and only if there exist qudit states $\ket{\alpha_k}=\sum_{j=1}^d x_{k,j}\ket{j}$ such that
\begin{eqnarray*}
\rho&=&\sum_{k} \int(U\otimes U)\alpha_k^{\otimes 2}(U\otimes U)^{\dag} dU\\
    &=&\sum_{k} |x_{k,i}|^2|x_{k,j}|^2\op{\psi_{i,j}}{\psi_{i,j}}.\\
\Leftrightarrow \chi:&=&(\chi_{ij})_{d\times d}=\sum_{k} \vec{x_k}\vec{x_k}^{T}.
\end{eqnarray*}
where $dU$ ranging over all diagonal unitaries, and $\vec{x_k}=(|x_{k,1}|^2,\cdots,|x_{k,d}|^2)^{T}\in\mathbb{R}^d$.

Recall that the cone of complete completely positive matrices \cite{DIA61,GW80} is define as
\begin{eqnarray*}
\mathcal{C}=\{\sum_{i}\vec{y_k}\vec{y_k}^{T}:\vec{y_k}\in\mathbb{R}^d_{+}\},
\end{eqnarray*}
where $\mathbb{R}^d_{+}$ stands for the $d$-dimensional vector space whose entries are all non-negative.

It is widely known that the decision problem on checking the completely positivity of given matrix is NP-Hard for general $d$ while for $d=3,4$, checking that the matrix is positive semidefinite and has all entries $\geq 0$ is both necessary and sufficient. Formally,
\begin{theorem}
It is NP-Hard to decide whether a given $d\otimes d$ GDS state is separable. On the other hand, $\rho=\sum_{i,j=1}^d\chi_{i,j} \op{\psi_{i,j}}{\psi_{i,j}}$ is separable if and only if $\chi=(\chi_{ij})_{d\times d}$ is semi-definite positive.
\end{theorem}
In other words, we have the following: PPT criterion is not sufficient for detect the entanglement for bosonic states, even for GDS states unless at least $P=NP$, which is highly impossible.

\textit{Conclusion---}In this paper, we study the separability problem of bosonic system. An analytical condition for the separability of $n$-qubit states whose eigenstates are Dicke states is demonstrated. For bipartite qudit system, we present a class of standard bosonic states, for general $d$, its separability is NP-hard, while for $d=3,4$, the condition of separability is provided.

It is still not clear that whether there exist easily verified analytical conditions for the separability of general $n-$qubit bosonic states?

We thank Prof. John Watrous, Prof. Debbie Leung and Prof. Bei Zeng for their comments.
NY is supported by NSERC, NSERC DAS, CRC, and CIFAR.

\end{document}